\documentclass[aps,twocolumn,prb,preprintnumbers,amsmath,amssymb]{revtex4-2}
\usepackage{graphicx}
\usepackage{dcolumn}
\usepackage{bm}
\usepackage{amsmath}
\usepackage{times}
\usepackage{natbib}

\begin{document}
\title{Photoinduced melting dynamics and collective mode in a correlated charge-ordered system}
\author{Yasuhiro Tanaka$^{1}$ and Hitoshi Seo$^{2}$}
\affiliation{$^1$Kanazawa Institute of Technology, Kanazawa 921-8501, Japan}
\affiliation{$^2$RIKEN Center for Emergent Matter Science, Wako 351-0198, Japan}
\date{\today}
\begin{abstract} 
We theoretically investigate the transient spectral function during the photoinduced melting of charge order in a correlated electron system, to unravel the dynamical processes triggered by different initial excitations. We employ a one-dimensional interacting spinless fermion model introducing a pulsed laser light, and perform a comparative study by the Hartree-Fock approximation and by the exact diagonalization method to numerically solve the time-dependent Schr\"odinger equation. We find characteristic behavior in the transient spectral function, whose features  strongly depend on the pump light frequency $\omega_p$. When $\omega_p$ is resonant with the collective phase mode of frequency $\Omega_c\simeq \Delta_{\rm CO}/2$, where $\Delta_{\rm CO}$ is the charge gap, the transient spectral function exhibits a photoinduced in-gap weight which triggers large responses. With increasing the laser intensity, the development of in-gap weight directly turns into the collapse of the gap. This charge-order destabilization process is in sharp contrast to the case of $\omega_p>\Delta_{\rm CO}$, where the photoirradiation induces interband electron-hole excitations giving rise to a shrinkage of the gap. The impact of quantum fluctuations and spatial inhomogeneity on the photoinduced dynamics is also discussed.
\end{abstract} 

\maketitle

\section{Introduction}
Photoinduced phase transitions offer a fascinating research field in condensed matter physics, 
where dramatic changes of electronic properties in solids appear upon stimuli of intense laser light
~\cite{Kirilyuk_RMP2010,Nasu_book2004,Koshihara_PhysRep2022}. 
Recent technological advancement has made it possible to manipulate materials in ultrashort time scale~\cite{Iwai_Crystals2012,Orenstein_PhysToday2012,Kampfrath_NatPhoto2013}, which has brought about discoveries of intriguing photoinduced phenomena and promotion of further theoretical studies in 
nonequilibrium physics~\cite{delaTorre_RMP2021,Aoki_RMP2014,Oka_AnnuRev2019}. In strongly correlated electron systems, a variety of photoinduced phase transitions, among which are insulator-to-metal transitions~\cite{Yonemitsu_Crystals2012,Iwai_JPSJ2006,Fiebig_ApplPhysB2000,Kawakami_PRL2010,Iwai_PRL2003,Iwai_PRL2007}, magnetic transitions~\cite{Sato_AccChemRes2003,Ishihara_JPSJ2019}, and creation of hidden states that 
 emerge only in nonequilibrium condition~\cite{Tokura_JPSJ2006,Stojchevska_Science2014,Ichikawa_NatMater2011,Fausti_Science2011},  have been reported and intensively studied so far. 

When an interacting electron system is photoexcited, it may exhibit a unique photoinduced dynamics originating from a collective mode that is inherent in the system. In fact, observations of such a photoinduced collective mode have been widely reported in, for 
 instance, systems with charge density wave~\cite{Schaefer_PRB2014,Ishikawa_Science2015}, superconductivity~\cite{Matsunaga_Science2014,Tsuji_PRB2015,Sun_PRR2020}, and excitonic order~\cite{Murakami_PRB2020,Bretscher_SciAdv2021,Tanaka_PRB2018,Tanabe_PRB2021}. In these systems, a coherent oscillation 
 of the underlying order parameter gives rise to a peculiar photoresponse in physical quantities, which is distinct from noninteracting 
 systems where the photoresponse basically originates from single-particle excitations. Although the photoexcited collective modes
have been identified in various materials, their role in photoinduced phase transitions where an initial ordered state 
is converted into a disordered state is elusive. 

To this end, we have conducted time-dependent Hartree-Fock (HF) simulations for a charge-order (CO) system in one dimension, which indicate that the collective mode indeed plays a key role in the CO melting dynamics and thus the photoinduced phase transition exhibits a strong dependence on the frequency of the incident light~\cite{Seo_PRB2018,Seo_PRB2024}. 
When the light frequency $\omega_p$ is larger than the charge gap $\Delta_{\rm CO}$, photoexcitation creates
photocarriers in the conduction band~\cite{Golez_PRB2016,Murakami_CommPhys2022} leading to the suppression of the CO. On the other hand, when $\omega_p$ is resonant with the collective mode, whose energy $\Omega_c$ is near a half of $\Delta_{\rm CO}$ ($\Omega_c\simeq \Delta_{\rm CO}/2$), a large photoresponse and the efficient CO destabilization appear. We note that the destabilization process also depends much on the light pulse width \cite{Seo_PRB2024}. The use of ultrashort pulses enables us to realize the collective mode excitations even when $\omega_p$ is off-resonant, which results in expanding the range of $\omega_p$ where the efficient CO collapse occurs.

These results raise us a question about how the emergent collective mode leads to the photoinduced phase transitions. Specifically, when we use a moderate pulse width for which the effective frequency range is sufficiently narrow, the CO destabilization process for $\omega_p<\Delta_{\rm CO}$ should be distinct from that with $\omega_p>\Delta_{\rm CO}$. This is because photocarrier generation due to 
direct excitation of electron-hole pairs, which occurs in the latter, is ineffective in the former. 
It is also important to clarify how quantum fluctuations affect the dynamics, which is beyond the scope of the HF theory. 

In this paper, we study the CO melting dynamics using a one-dimensional spinless fermion model with the nearest neighbor 
Coulomb interaction that induces a CO insulating ground state. We simulate time evolution of the system exposed to a pulsed electric field of light by two methods: the HF approximation and exact diagonalization (ED)~\cite{Takahashi_PRL2002,Maeshima_JPSJ2005,Miyashita_JPSJ2010,Lu_PRL2012,Hashimoto_JPSJ2014,Hashimoto_JPSJ2015}. This enables us to examine the validity and limitation of the HF approach on the photoinduced dynamics. The time-dependent Schr\"odinger equation is numerically solved and transient spectral functions are computed, which provide direct access to the mechanism of the initial processes of  photo-induced phase transitions. 
We show that for $\omega_p\simeq \Omega_c$, an in-gap weight is induced by the laser light in the transient spectral functions indicating the energy absorption due to excitation of the collective mode. With increasing the 
pump strength, the gapped structure is destroyed by the development of this in-gap weight. This behavior is markedly different from the case with $\omega_p>\Delta_{\rm CO}$ where interband excitations are seen and lead to the reduction of the gap magnitude. We also discuss the effects of quantum fluctuations and inhomogeneity embedded in a bulk system~\cite{Seo_PRB2018,Seo_PRB2024,Picano_PRB2023} on CO melting processes. 

\section{Model and Method}
\subsection{Model}
We consider a model of interacting spinless fermions in a one-dimensional lattice irradiated by a laser light, which is given by the Hamiltonian,
\begin{equation}
\mathcal{H}=\sum_{i}\left(te^{iA(\tau)}\ c_{i}^{\dagger}c_{i+1}+\text{H.c.}\right) + V\sum_{i} n_{i} n_{i+1}, 
\label{eqn1}
\end{equation}
where $c_{i} (c_{i}^{\dagger})$ is the annihilation (creation) operator for a spinless fermion at the $i$-th site,
and $n_{i} \equiv c_{i}^{\dagger}c_{i}$ is the fermion number operator. We write the transfer integral between nearest neighbor sites as 
$t$ and the nearest neighbor Coulomb repulsion as $V$. The effect of photoexcitation is introduced as a time ($\tau$) dependent gauge field $A(\tau)$ for the electric field of laser light in the Peierls phase of fermion hopping terms. 
We assume a Gaussian pulsed laser centered at $\tau=0$, for which $A(\tau)$ is written as
\begin{align}
A(\tau)=\frac{A_{p}}{\sqrt{2\pi}\tau_{p}}\exp{\left(-\frac{\tau^{2}}{2\tau_{p}^{2}}\right)}\cos(\omega_{p}\tau),
\label{eqn2}
\end{align}
where $A_{p}$, $\omega_p$, and $\tau_{p}$ are the strength, frequency,  and pulse width of the 
pump light, respectively. The electron density is fixed at half-filling: 
we consider $N_e/N=1/2$ where $N_e$ ($N$) is the total number of fermions (sites). 
The lattice constant, the light velocity, the elementary charge, and $\hbar$ 
are taken as unity. 
In the following, we set $t=1$ as the unit of energy (and time $1/t$), $V=3$ where the system shows a stable CO ground state, and $\tau_p=3$. 
 
\subsection{Time-dependent Hartee-Fock approximation}
In the HF theory, we approximate the Coulomb interaction term in Eq. (\ref{eqn1}) 
as $n_in_{i+1}\rightarrow n_i\langle n_{i+1}\rangle+n_{i+1}\langle n_i\rangle-\langle n_i\rangle\langle n_{i+1}\rangle
-\langle c_i^{\dagger}c_{i+1}\rangle c_{i+1}^{\dagger}c_i-\langle c_{i+1}^{\dagger}c_{i}\rangle c_{i}^{\dagger}c_{i+1}
+\langle c_i^{\dagger}c_{i+1}\rangle \langle c_{i+1}^{\dagger}c_i\rangle$. 
We consider the CO with twofold periodicity for which the order parameter $\phi$ is defined as
\begin{equation}
\phi =\frac{(-1)^i}{2}(\langle n_{i+1}\rangle-\langle n_{i}\rangle).
\label{eqn3}
\end{equation}
After determining the HF self-consistent solution in the ground state ($A_p=0$), we calculate photoinduced dynamics by numerically solving the time-dependent 
Schr\"odinger equation~\cite{Terai_PTP1993,Kuwabara_JPSJ1995,Miyashita_JPSJ2003,Tanaka_JPSJ2010} 
that is discretized as
\begin{equation}
|\psi_{k,\nu}(\tau+\Delta \tau)\rangle={\rm exp}\Bigl[-i
\Delta \tau {\mathcal H}^{\rm HF}_{k}\Bigl(\tau+\frac{\Delta \tau}{2}\Bigr)\Bigr]|\psi_{k,\nu}(\tau)\rangle ,
\label{eqn4}
\end{equation}
where ${\mathcal H}^{\rm HF}_{k}(\tau)$ is the HF Hamiltonian in momentum space at time $\tau$, $|\psi_{k,\nu}(\tau)\rangle$ the $\nu$-th ($\nu=1,2$) one-particle state with wave number $k$ at time $\tau$. In the calculation of Eq. (\ref{eqn4}), we employ a numerical method which gives $|\psi_{k,\nu}(\tau+\Delta \tau)\rangle$ with errors of the order of $(\Delta \tau)^3$~\cite{Terai_PTP1993,Kuwabara_JPSJ1995,Miyashita_JPSJ2003}, and use $\Delta \tau =0.01$ which guarantees sufficient numerical accuracy. 

\subsection{Many-body dynamics with exact diagonalization}
The many-body effect that is ignored in the HF theory is examined by calculating the time evolution of the exact ground-state wave function obtained by the ED method. To describe the ground state with a long-range CO in finite systems, we add a small staggered potential to Eq. (\ref{eqn1}), which is written as
\begin{equation}
{\mathcal H}_{\rm sp}=\delta_{\rm sp}\sum_i n_i(-1)^i,
\label{eqn5}
\end{equation}
and diagonalize ${\mathcal H}_{\rm tot}={\mathcal H}+{\mathcal H}_{\rm sp}$ using the Lanczos method where the periodic boundary condition is applied. We consider systems with $N=18$ and $22$ for which the closed shell is realized at half-filling. The time evolution of the system under pulsed laser light is simulated by numerically solving the time-dependent Schr\"odinger equation~\cite{Takahashi_PRL2002,Maeshima_JPSJ2005,Lu_PRL2012,Hashimoto_JPSJ2014,Hashimoto_JPSJ2015,Miyashita_JPSJ2010}, 
\begin{equation}
|\Psi(\tau+\Delta \tau)\rangle =\exp\Bigl[-i\Delta \tau{\mathcal H}_{\rm tot}\Bigl(\tau+\frac{\Delta \tau}{2}\Bigr)\Bigr]|\Psi(\tau)\rangle, 
\label{eqn6}
\end{equation}
where $|\Psi(\tau)\rangle$ is the many-body wave function at time $\tau$. We use $\Delta \tau=0.01$ and $\delta_{\rm sp}=0.03$. 

\subsection{Transient spectral functions}
The microscopic mechanism of the photoinduced destabilization of CO is investigated by calculating transient spectral functions which are given by~\cite{Freericks_PRL2009,Sentef_NatC2015}
\begin{eqnarray}
A_k(\varepsilon, \tau_{\rm pr})&=&A^<_k(\varepsilon, \tau_{\rm pr})+A^>_k(\varepsilon, \tau_{\rm pr}), \\
\label{eqn7}
A^<_k(\varepsilon, \tau_{\rm pr})&=&{\rm Im}\sum_{\alpha}\int d\tau_1 d\tau_2
s(\tau_1-\tau_{\rm pr})s(\tau_2-\tau_{\rm pr}) \nonumber \\
\label{eqn8}
&&\times e^{i\varepsilon (\tau_1-\tau_2)}G^{<}_{k,\alpha\alpha}(\tau_1,\tau_2),\\
A^>_k(\varepsilon, \tau_{\rm pr})&=&-{\rm Im}\sum_{\alpha}\int d\tau_1 d\tau_2
s(\tau_1-\tau_{\rm pr})s(\tau_2-\tau_{\rm pr}) \nonumber \\
&&\times e^{i\varepsilon (\tau_1-\tau_2)}G^{>}_{k,\alpha\alpha}(\tau_1,\tau_2), 
\label{eqn9}
\end{eqnarray}
where $A^<_k(\varepsilon, \tau_{\rm pr})$ and $A^>_k(\varepsilon, \tau_{\rm pr})$ 
correspond to photoemission and inverse photoemission spectra, respectively, and $\tau_{\rm pr}$ is the probe time at which we ``measure'' the spectra.
Here $G^{<}_{k,\alpha\beta}(\tau_1,\tau_2)$ and $G^{>}_{k,\alpha\beta}(\tau_1,\tau_2)$ are the lesser and greater non-equilibrium Green's functions 
which are defined as
\begin{eqnarray}
G^{<}_{k,\alpha\beta}(\tau_1,\tau_2)&=&i\langle c^{\dagger}_{k,\beta}
(\tau_2)c_{k,\alpha}(\tau_1)\rangle, \\
\label{eqn10}
G^{>}_{k,\alpha\beta}(\tau_1,\tau_2)&=&
-i\langle c_{k,\alpha}(\tau_1)c^{\dagger}_{k,\beta}(\tau_2)\rangle ,
\label{eqn11}
\end{eqnarray}
where the operator $c^{\dagger}_{k,\alpha}$ ($c_{k,\alpha}$) is the Fourier transform 
of $c^{\dagger}_{\mu, \alpha}$ ($c_{\mu, \alpha}$) with $\mu$ and $\alpha$ ($=1,2$) being the index for the unit cell and that for the sublattice in the unit cell, respectively. We assume the Gaussian function for the probe pulse, 
\begin{eqnarray}
s(\tau)=\frac{1}{\sqrt{2\pi}\sigma_{\rm pr}}
\exp \left(-\frac{\tau^2}{2\sigma_{\rm pr}^2}\right),
\label{eqn12}
\end{eqnarray}
with the pulse width $\sigma_{\rm pr}=5$. The time integrations in Eqs. (\ref{eqn8}) and (\ref{eqn9}) are performed in the range of 
$\tau_{\rm pr}-4\sigma_{\rm pr}<\tau_1,\tau_2<\tau_{\rm pr}+4\sigma_{\rm pr}$. In the following, we set $\tau_{\rm pr}=10$ unless otherwise noted.

\section{Results}
\subsection{Time-dependent Hartree-Fock approximation}
\begin{figure}[]
\includegraphics[width=9cm,clip]{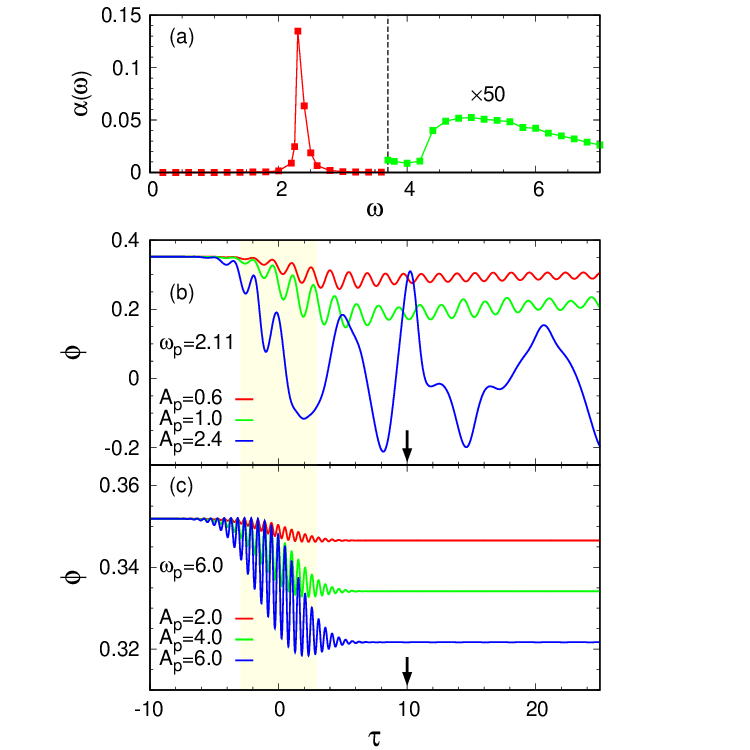}
\caption{(a) Linear optical absorption spectrum $\alpha(\omega)$. Time evolution of $\phi$ under pulsed laser light with  (b) $\omega_p=2.11$ and (c) $\omega_p=6.0$ for $\tau_p=3$. The shaded region indicates the time domain where the pump light is applied. The arrows mark the probe time $\tau=\tau_{\rm pr}$ of the transient spectral functions. We use $V=3$ and $N=400$.}
\label{fig1}
\end{figure}
For $V=3$, the HF theory gives the CO ground state with the charge gap $\Delta_{\rm CO}=4.22$. We use $N=400$ for which the finite 
size effect is negligible. In Fig. \ref{fig1}(a), we show the linear optical absorption spectrum $\alpha(\omega)$ which is obtained by calculating the 
increment in the total energy under a weak continuous-wave field with damping~\cite{Tanaka_JPSJ2010}. 
Specifically, we replace $A(\tau)$ in Eq. (2) by 
\begin{equation}
A(\tau)=A_{\rm cw} \theta(\tau)e^{-\gamma \tau}\sin \omega \tau,
\label{eqn13}
\end{equation}
with $A_{\rm cw}=0.01$ and $\gamma=0.02$, and compute the increase in total energy at sufficiently large $\tau$. We note that $\alpha(\omega)$ is directly related to the regular part of optical conductivity $\sigma_{\rm reg}(\omega)$, which we discuss in Appendix A. Because of the existence of the collective mode, $\alpha(\omega)$ has a sharp peak at $\omega=\Omega_c=2.3$ ($\simeq \Delta_{\rm CO}/2$)~\cite{Seo_PRB2018}. For $\omega>\Delta_{\rm CO}$, a broad absorption band appears, which is due to interband single-particle excitations. 

Then, in Fig. \ref{fig1}(b), we show time evolution of $\phi$ upon the irradiation of the pulsed laser light with $\omega_p=2.11=\Delta_{\rm CO}/2$. As discussed in our previous works \cite{Seo_PRB2018,Seo_PRB2024}, we find a large response with an oscillation  indicating the photoinduced collective mode. The oscillation persists even after photoexcitation and its amplitude increases with increasing $A_p$. For $A_p=2.4$, $\phi$ reaches zero; a momentary disappearance of the CO occurs~\cite{Seo_PRB2018}. On the other hand, as seen in Fig. \ref{fig1}(c), the time evolution of $\phi$ when we use the light with $\omega_p=6.0$ ($>\Delta_{\rm CO}$) indicates that, for the excitation across the gap, the photoresponse mainly appears in the time domain where the light pulse is applied. Compared to the case with $\omega_p=2.11$, the decrease in $\phi$ is much smaller and $\phi$ does not reach zero even at $A_p=6.0$. 

\begin{figure}[]
\includegraphics[width=9.0cm]{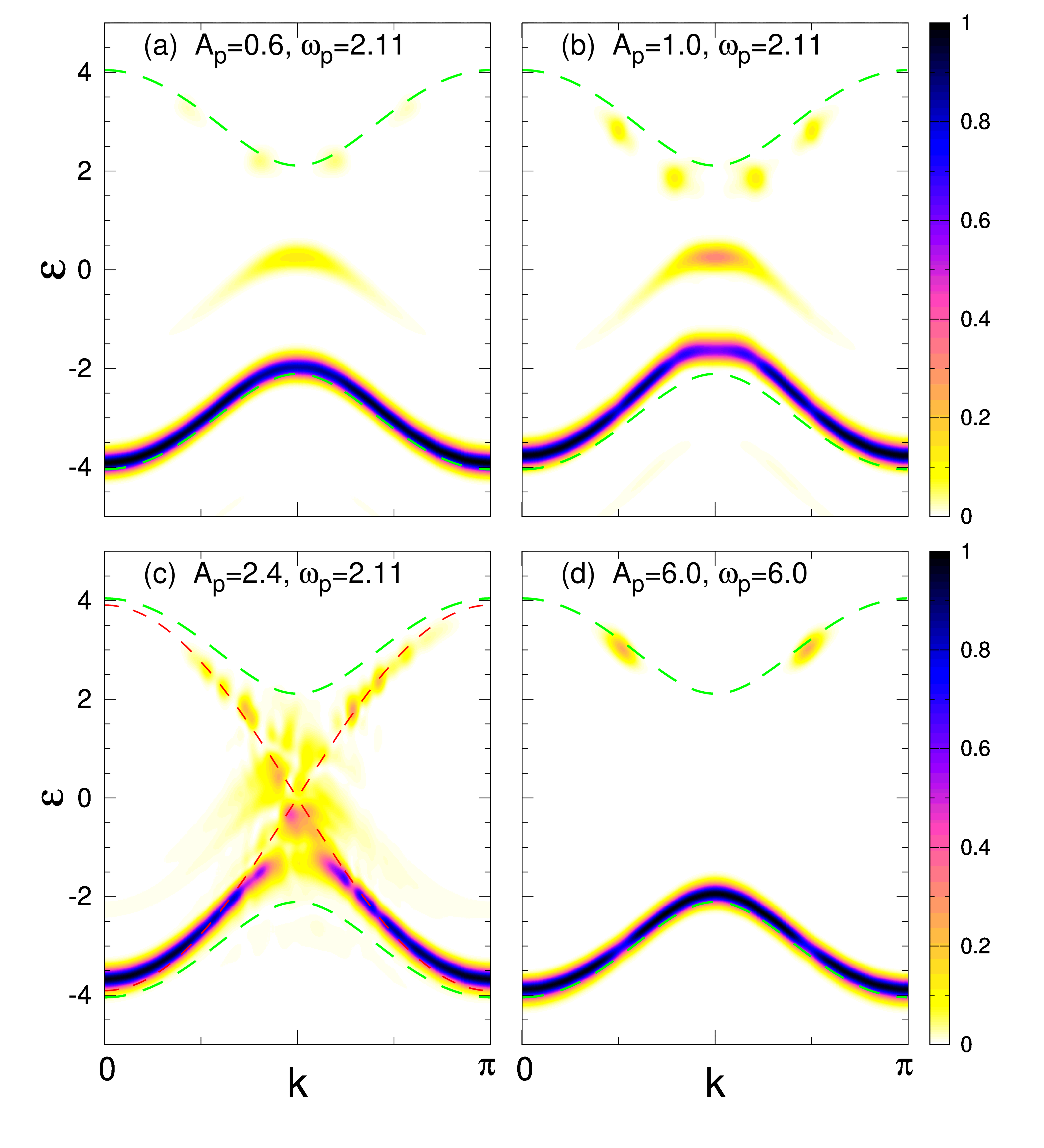}
\caption{Photoemission spectra $A^<_k(\varepsilon,\tau_{\rm pr})$ for (a) $A_p=0.6$, (b) $1.0$, and (c) $2.4$ with $\omega_p=2.11$. (d) $A^<_k(\varepsilon,\tau_{\rm pr})$ for $A_p=6.0$ with $\omega_p=6.0$. The green dashed curves indicate the HF bands of the CO ground state. In (c), the red dashed curves show the HF metallic bands with $\phi=0$.}
\label{fig2}
\end{figure}

Now let us compare the transient spectra in the two cases above. In Figs. \ref{fig2}(a)--(c), we show $A_k^<(\varepsilon,\tau_{\rm pr})$ with $\omega_p=2.11$ for different values of $A_p$. Hereafter, the origin of $\varepsilon$ is shifted by $V$ so that the midpoint of the initial gap is located at $\varepsilon=0$. 
Importantly, for $A_p=0.6$, one can clearly see that an in-gap weight appears at $\varepsilon\simeq 0$ and $k\simeq \pi/2$ with a dispersion similar to the ground-state occupied band [Fig. \ref{fig2}(a)]. Besides the in-gap weight, there are weights due to interband charge excitations, appearing at discrete $k$-points in the originally unoccupied bands at 
 $\varepsilon \simeq 2.1$ and $\varepsilon \simeq 3.2$, indicating absorption of two and three photons, respectively. Since the photocarrier generation does not occur by a single photon absorption for $\omega_p<\Delta_{\rm CO}$, the in-gap weight manifests a photoexcited collective mode of energy $\Omega_c\simeq \Delta_{\rm CO}/2$. With increasing $A_p$, the transient valence band shifts upward as shown in Fig. \ref{fig2}(b), reflecting the decrease in $\phi$ [Fig. \ref{fig1}(b)]. In addition, intensities of the in-gap weight and the interband charge excitations increase. By increasing the intensity furthermore, for $A_p=2.4$ [Fig. \ref{fig2}(c)], these photoinduced weights and the shifted valence band merge with each other near $k\simeq \pi/2$, and almost lay on the HF metallic bands ($\phi=0$) calculated in the ground state. This drastic reconstruction of spectral function corresponds to the destabilization and momentary melting of the CO shown in Fig. \ref{fig1}(b). 
 
 In clear contrast, in Fig. \ref{fig2}(d), we show $A_k^<(\varepsilon,\tau_{\rm pr})$ with $\omega_p=6.0$ and $A_p=6.0$ where the weights due to  interband excitations by a single photon absorption are seen at $\varepsilon \simeq 3.0$. The in-gap weight seen when photoexciting the collective mode [Figs. 2(a)--2(c)] is totally absent and the valence band shift is very small, which is consistent with the small responses seen in the time evolution of $\phi$ shown in Fig. \ref{fig1}(c). 


\begin{figure}
\begin{center}
\includegraphics[width=9cm]{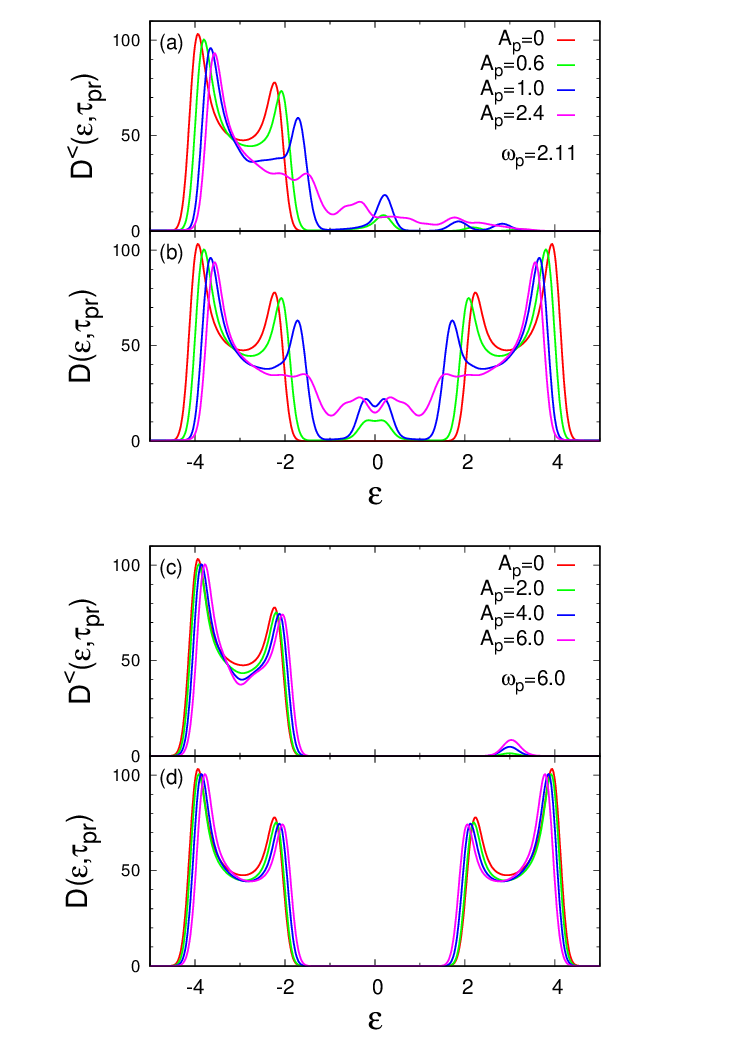}
\end{center}
\caption{Momentum integrated DOS (a) $D^<(\varepsilon,\tau_{\rm pr})$ and (b) $D(\varepsilon,\tau_{\rm pr})$ for different values of $A_p$ with $\omega_p=2.11$. (c) $D^<(\varepsilon,\tau_{\rm pr})$ and (d) $D(\varepsilon,\tau_{\rm pr})$ with $\omega_p=6.0$. We use $\tau_{\rm pr}=10$.}
\label{fig3}
\end{figure}

To see the evolution of the weight transfer and the overall gap structure, in Figs. \ref{fig3}(a) and \ref{fig3}(b), we show the $k$-integrated density of states (DOS) $D^<(\varepsilon,\tau_{\rm pr})=\sum_k A^<_k(\varepsilon,\tau_{\rm pr})$ and $D(\varepsilon,\tau_{\rm pr})=\sum_k A_k(\varepsilon,\tau_{\rm pr})$, respectively, for $\omega_p=2.11$ corresponding to the cases shown in Figs. \ref{fig1}(b) and \ref{fig2}(a)--\ref{fig2}(c), whereas those for $\omega_p=6.0$ are shown in Figs. \ref{fig3}(c) and \ref{fig3}(d), corresponding to Figs. \ref{fig1}(c) and \ref{fig2}(d). For the former situation, the weight at the top of the valence band is transferred to the in-gap weight at $\varepsilon \simeq 0$ with increasing $A_p$ as seen in Figs. \ref{fig3}(a) and \ref{fig3}(b). Notably, the development of the in-gap weight collapses the gap structure from inside the original gap. On the other hand, for latter cases, the gap only slightly shrinks by the interband excitations as shown in Figs. \ref{fig3}(c) and \ref{fig3}(d).

\begin{figure}[]
\begin{center}
\includegraphics[width=9cm]{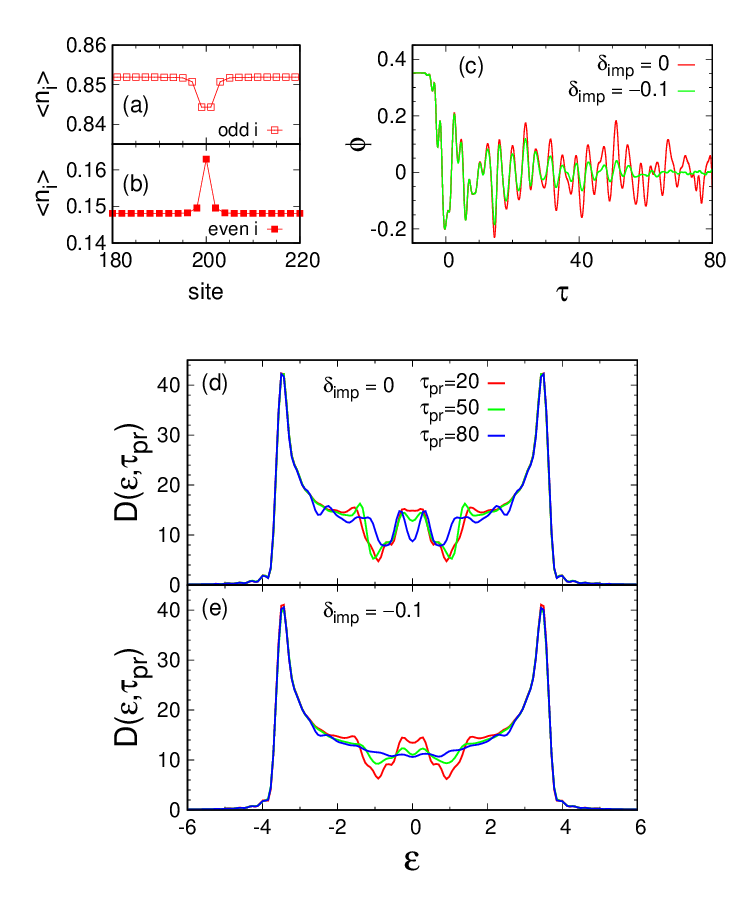}
\end{center}
\caption{Electron densities $\langle n_i\rangle$ near the impurity site $i=i_p$ for (a)
charge-rich and (b) charge-poor sites in the ground state with $\delta_{\rm imp}=-0.1$. 
(c) Time evolutions of $\phi$ with $\delta_{\rm imp}=0$ and $-0.1$. $D(\varepsilon,\tau_{\rm pr})$ for different values of $\tau_{\rm pr}$ with (d) $\delta_{\rm imp}=0$ and (e) $\delta_{\rm imp}=-0.1$. In (c)-(e), we use $\omega_p=2.11$, $\tau_p=3$, and $A_p=2.4$.}
\label{fig4}
\end{figure}

We note that the results presented here do not qualitatively change when we take longer widths ($\tau_p>3$).
On the other hand, when we use shorter pulses,  the frequency range of the light expands and the momentary CO disappearance can occur even at $\omega_p\simeq \Delta_{\rm CO}$ \cite{Seo_PRB2024}. We will discuss the pulse width dependence in our future work \cite{Tanaka_XXX2025}.

Finally, we discuss the results at a longer time range, where the CO melting process is affected by inhomogeneity. As seen in Fig. \ref{fig1}(b), for $\omega_p=2.11$ and large $A_p$, $\phi$ keeps oscillating largely after the photoexcitation and the CO disappears when $\phi$ changes its sign. As we have shown in previous studies~\cite{Seo_PRB2018,Seo_PRB2024}, an inhomogeneity embedded in the bulk CO serves as a source of relaxation, by which the system reaches a photoinduced metallic state with a persistent absence of the long-ranged CO. To elucidate such dynamics, we examine a long-time behavior of $D(\varepsilon,\tau_{\rm pr})$ by varying $\tau_{\rm pr}$. We introduce a single impurity, the effect of which is incorporated by adding an on-site potential at site $i_{p}$, 
\begin{equation}
{\mathcal H}_{\rm imp}=\delta_{\rm imp} n_{i_p},
\label{eqn14}
\end{equation}
to Eq. (1). We set $\delta_{\rm imp}=-0.1$ and $i_p=N/2$, and perform time-dependent HF calculations in the real space representation. In Figs. \ref{fig4}(a) and \ref{fig4}(b), we show electron densities $\langle n_i\rangle$ near $i=i_p$ in the ground state for charge-rich and charge-poor sites, respectively: the impurity potential disturbs the bulk electron distribution near $i=i_p$. In Fig. \ref{fig4}(c), time evolultions of $\phi$ with and without the impurity potential are shown for $\omega_p=2.11$ and $A_p=2.4$; for the case with the impurity, the average over the whole system is plotted. For $\delta_{\rm imp}=0$, $\phi$ oscillates persistently, whereas it relaxes to zero for $\delta_{\rm imp}=-0.1$, as discussed in Refs. \cite{Seo_PRB2018,Seo_PRB2024}. 

In Figs. \ref{fig4}(d) and \ref{fig4}(e), we show transient DOS for the whole system $D(\varepsilon,\tau_{\rm pr})$ for different values of $\tau_{\rm pr}$ without the impurity and that with $\delta_{\rm imp}=-0.1$, respectively.  For $\delta_{\rm imp}=0$, $D(\varepsilon,\tau_{\rm pr})$ exhibits time variation even at $\tau_{\rm pr}=80$ especially in the region near $\varepsilon=0$, which corresponds to the large oscillation in $\phi$ after the photoexcitation seen in Fig. \ref{fig1}(b). On the other hand, for $\delta_{\rm imp}=-0.1$, $D(\varepsilon,\tau_{\rm pr})$ shows the relaxation and a metallic-like DOS is realized at $\tau_{\rm pr}=80$. This directly shows that the system has almost totally loses its CO nature and reaches to a photoinduced metallic phase, as was anticipated from the simulations in Refs. \cite{Seo_PRB2018,Seo_PRB2024}.

\begin{figure}[]
\begin{center}
\includegraphics[width=9.5cm]{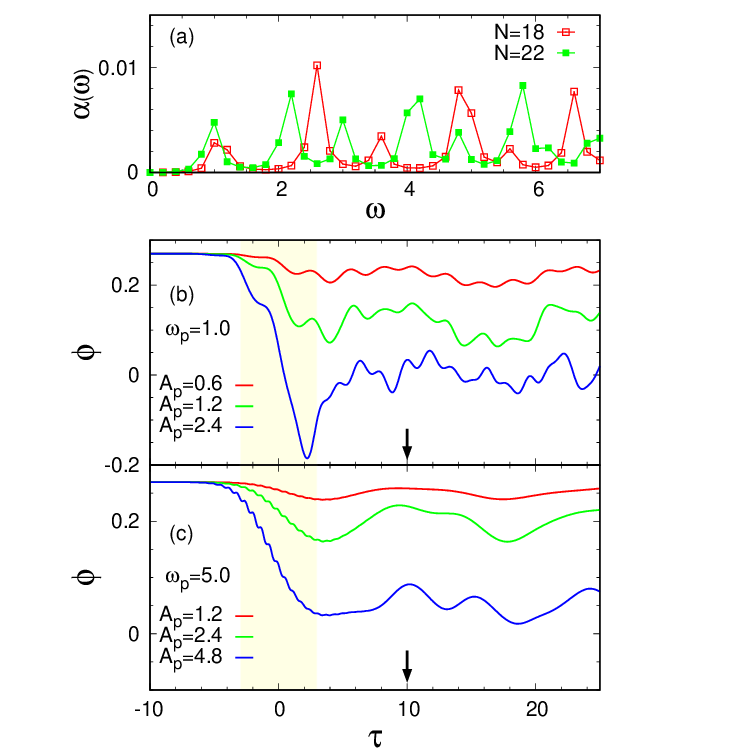}
\end{center}
\caption{(a) Linear optical absorption spectra $\alpha(\omega)$ for $N=18$ and $22$. Time evolution of $\phi$ under pulsed laser light  
with (b) $\omega_p=1.0$ and (c) $\omega_p=5.0$ for $\tau_p=3$ and $N=22$. 
The shaded region indicates the time domain where the pump light is applied. The arrows mark the probe time $\tau=\tau_{\rm pr}$ of the transient spectral functions.
}
\label{fig5}
\end{figure}

\subsection{Many-body dynamics with exact diagonalization}
In this section, we show results of the calculations using the ED method. We should note that, compared to the time-dependent HF method discussed above, quantum fluctuations can be fully treated but the cluster size one can consider due to numerical limitation is considerably smaller so that finite-size effects should be carefully characterized. The charge gap before light irradiation is defined as $\Delta_{\rm CO}=\mu_+-\mu_-$ with $\mu_+=E_0(N,N_e+1)-E_0(N,N_e)$ and $\mu_-=E_0(N,N_e)-E_0(N,N_e-1)$ where $E_0(N,N_e)$ is the ground 
state energy for a system of $N$ sites with $N_e$ electrons. For $V=3$ and at half filling $(N_e=N/2)$, we obtain $\Delta_{\rm CO}=1.77$ for $N=22$ and $\Delta_{\rm CO}=1.99$ for $N=18$. 

Figure \ref{fig5}(a) shows the linear optical absorption spectra $\alpha(\omega)$ calculated from the many-body dynamics for the two cluster sizes, where we use $A(\tau)$ in Eq. (\ref{eqn13}) with $A_{\rm cw}=0.01$ and $\gamma=0.02$. The spectra consist from discrete peaks, which is due to the finite size effect~\cite{Hashimoto_PRB2017,Ishihara_JPSJ97}. The peak at $\omega= \Omega_c=1.0$ (which is slightly larger than 
$\Delta_{\rm CO}/2$, but is actually half of the single-particle gap in $D(\varepsilon,\tau_{\rm pr}$) as we will see later) corresponds to the collective phase mode of the CO that has the lowest excitation energy of the system, as in the case of HF approximation shown in Fig. \ref{fig1}(a). In fact, by calculating up to the 3rd lowest eigenvalue, $E_2(N,N_e)$, we obtain 
the lowest excitation energy from doubly-degenerate ground states as $E_2(N,N_e)-E_0(N,N_e)=0.94$ for $N=22$ and 1.07 for $N=18$, which is consistent with the peak position $\omega= \Omega_c$ in $\alpha(\omega)$ and thus supports our assignment. We note that other peaks with $\omega>\Omega_c$ in $\alpha(\omega)$ might come from excitation of collective motions of CO with nonzero momenta~\cite{Hashimoto_PRB2017}. The finite size effect for the frequency $\Omega_c$ is small, whereas the other peak frequencies vary largely with $N$. 

Considering these results, we discuss the dynamics upon stimuli of a pulsed laser light with $\tau_p=3$. In Figs. \ref{fig5}(b) and \ref{fig5}(c), we show time evolution 
of $\phi$ with $\omega_p=1.0$ and $5.0$, respectively, for $N=22$. For the case of $\omega_p=1.0$, $\phi$ shows a large response with an oscillation which continues after the photoexcitation, and it reaches nearly zero for $A_p=2.4$. Compared to the results with the HF approximation shown in Fig. \ref{fig1}(b), a clear sinusoidal oscillation for small $A_p$ is not prominent in the ED results. As for the case of $\omega_p=5.0$ ($>\Delta_{\rm CO}$), the photoresponse is smaller than that for $\omega_p=1.0$. However, $\phi$ reaches nearly zero for large values of $A_p$, which is different from the HF results.

\begin{figure}[]
\begin{center}
\includegraphics[width=10cm]{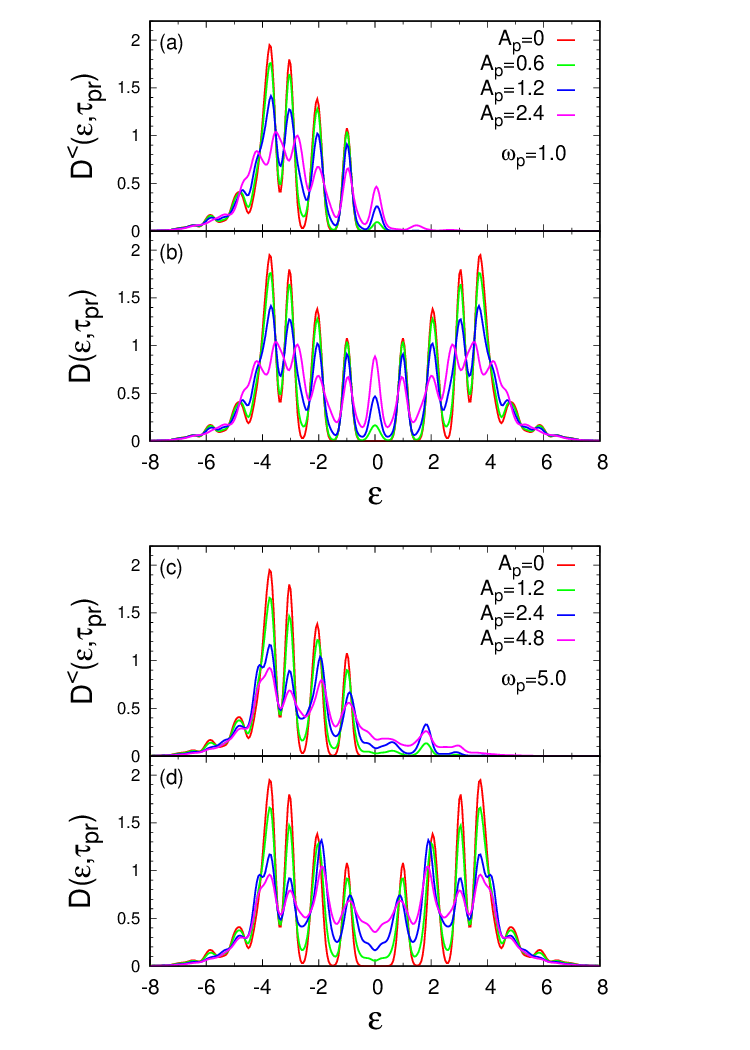}
\end{center}
\caption{Momentum integrated DOS (a) $D^<(\varepsilon,\tau_{\rm pr})$ and (b) $D(\varepsilon,\tau_{\rm pr})$ for different values of $A_p$ with $\omega_p=1.0$. (c) $D^<(\varepsilon,\tau_{\rm pr})$ and (d) $D(\varepsilon,\tau_{\rm pr})$ with $\omega_p=5.0$. We use $\tau_{\rm pr}=10$ and $N=18$.}
\label{fig6}
\end{figure}

\begin{figure}[t]
\includegraphics[width=7.0cm]{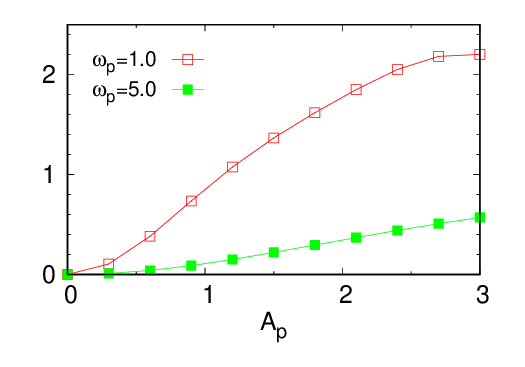}
\caption{In-gap DOS as a function of $A_p$ with $\omega_p=1.0$ and $5.0$.}
\label{fig7}
\end{figure}

In Figs. \ref{fig6}(a) and \ref{fig6}(b), we show $D^<(\varepsilon,\tau_{\rm pr})$ and $D(\varepsilon,\tau_{\rm pr})$ with $\omega_p=1.0$, respectively, for $\tau_{\rm pr}=10$. In the calculations of transient spectra, we use $N=18$ due to computational constraint. We note that the results for $N=14$ are qualitatively consistent with those for $N=18$ and therefore the finite size effect does not alter our discussions below. Without the light ($A_p=0$), $D(\varepsilon,\tau_{\rm pr})$ indicates a single-particle gap of about 2.0 [Fig. \ref{fig6}(b)], which is close to $\Delta_{\rm CO}$. For $A_p\neq 0$, there appears peak-like in-gap structure at $\varepsilon\simeq 0$ whose intensity increases with increasing $A_p$. For $A_p=2.4$, the gap in $D(\varepsilon,\tau_{\rm pr})$ is collapsed from inside the gap structure as shown in Fig. \ref{fig6}(b). This characteristic behavior is qualitatively consistent with those obtained by the HF approximation shown in Fig. \ref{fig3}. The notable difference between the ED and HF results is that spectral weights at low energies ($-4.0\lesssim \varepsilon \lesssim -3.0$) in $D^<(\varepsilon,\tau_{\rm pr})$ decrease more largely in the ED results, which may be due to the correlation effect~\cite{Okamoto_NJP2019,Innerberger_EPJP2020}. 

For $\omega_p=5.0$, we depict $D^<(\varepsilon,\tau_{\rm pr})$ and $D(\varepsilon,\tau_{\rm pr})$ in Figs. \ref{fig6}(c) and \ref{fig6}(d), respectively. For $A_p=1.2$, there is a small spectral weight at $\varepsilon=1.8$ in $D^<(\varepsilon,\tau_{\rm pr})$ indicating interband excitations, and its intensity increases with increasing $A_p$. For $A_p=4.8$, $D^<(\varepsilon,\tau_{\rm pr})$ has a broad $\varepsilon$ dependence due to the suppression of $\phi$. The gap structure in $D(\varepsilon,\tau_{\rm pr})$ shrinks and collapses from outside the gap structure with increasing $A_p$, which is in contrast to the case of $\omega_p=1.0$. We note that 
$\phi$ and $D(\varepsilon,\tau_{\rm pr})$ exhibit no relaxation after photoexcitation even when we consider an impurity potential (not shown). This is different from the HF results, whose reason we speculate is that the coherence length of the photoexcited state is longer than the system size. 
In Fig. \ref{fig7}, we show the $A_p$ dependence of an in-gap DOS which is computed by an integration of $D(\varepsilon,\tau_{\rm pr})$ within  $-0.1<\varepsilon<0.1$ for $\omega_p=1.0$ and $\omega_p=5.0$. It is evident that the in-gap DOS for $\omega_p=1.0$ is much larger than that for $\omega_p=5.0$. This is due to the in-gap weight at $\varepsilon=0$ coming from the photoinduced collective mode. These contrastive behaviors of  $D(\varepsilon,\tau_{\rm pr})$ depending on the value of $\omega_p$ are consistent with those obtained by the HF approximation. 

\begin{figure}[t]
\includegraphics[width=9cm]{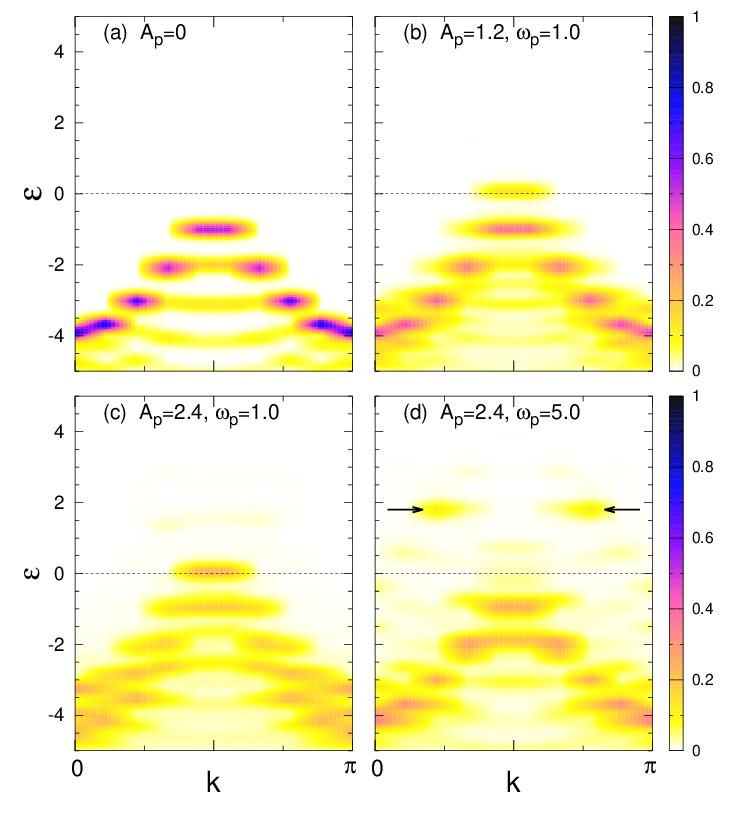}
\caption{Momentum dependence of $A^<_k(\varepsilon,\tau_{\rm pr})$ for (a) $A_p=0$, (b) $A_p=1.2$ and $\omega_p=1.0$, (c) $A_p=2.4$ and $\omega_p=1.0$, and (d) $A_p=2.4$ and $\omega_p=5.0$. Other parameters are $\tau_{\rm pr}=10$, $\tau_p=3$, and $N=18$. The dashed line in each panel indicates $\varepsilon =0$. In (d), the arrows indicate the transferred spectral weights at $\varepsilon =1.8$ and $k=2\pi/9, 7\pi/9$.}
\label{fig8}
\end{figure}

To examine the results in more detail, we show the momentum dependence of $A_k^<(\varepsilon,\tau_{\rm pr})$ for different $\omega_p$ and $A_p$ in Fig. \ref{fig8}. For $A_p=0$ (no photo-irradiation), a large peak at each $k$ forms the valence band of the CO ground state [Fig. \ref{fig8}(a)]. For $A_p=1.2$ and $\omega_p=1.0$, the in-gap weights appear near $k\simeq \pi/2$ ($4\pi/9$ and $5\pi/9$) and the valence band peaks are blurred due to photoexcitation [Fig. \ref{fig8}(b)]. For $A_p=2.4$, the in-gap weights further develop and they are connected to peaks forming a dispersion similar to the valence band, shifted to higher energy, as shown in Fig. \ref{fig8}(c). This is consistent with the HF results for $\omega_p\simeq \Omega_c$ in Figs. \ref{fig2}(a) and \ref{fig2}(b). For $A_p=2.4$ and $\omega_p=5.0$, spectral peaks in the ground state decrease and there is a spectral transfer to $\varepsilon =1.8$ at $k=2\pi/9$, $7\pi/9$ without notable dispersion around them, as shown in Fig. \ref{fig8}(d). This indicates the interband excitations as in the HF results shown in Fig. \ref{fig2}(d).

\begin{figure}[]
\begin{center}
\includegraphics[width=8cm]{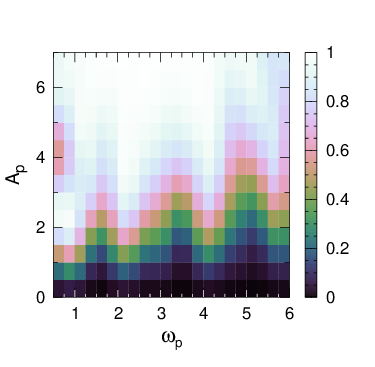}
\end{center}
\caption{Two-dimensional plot of $(\phi_0-|\phi_{\rm av}|)/\phi_0$ on the $\{\omega_p, A_p\}$ plane. We use $\tau_p=3$ and $N=22$.
}
\label{fig9}
\end{figure}

Finally, we discuss the decrease in $\phi$ due to photoexcitation by varying $\omega_p$ and $A_p$~\cite{Seo_PRB2018,Seo_PRB2024}. In Fig. \ref{fig9}, we show the two-dimensional plot of $(\phi_0-|\phi_{\rm av}|)/\phi_0$ on the $\{\omega_p, A_p\}$ plane for $\tau_p=3$ and $N=22$, where $\phi_0$ and $\phi_{\rm av}$ are the CO order parameter in the ground state and the time-averaged $\phi$ in the time domain of $10<\tau<30$, respectively. The structure of the two-dimensional plot can be understood essentially from the behavior of $\alpha(\omega)$ shown in Fig. \ref{fig5}(a); for $\omega_p$ with large $\alpha(\omega)$, $\phi$ efficiently decreases with small $A_p$, whereas for $\omega_p$ with small $\alpha(\omega)$, a large $A_p$ is needed to suppress $\phi$. In Appendix B, we show the two-dimensional plot with longer pulse width of $\tau_p=10$ for which the correspondence with $\alpha(\omega)$ is more apparent. In the HF results~\cite{Seo_PRB2018,Seo_PRB2024}, the decrease in $\phi$ is basically small when $\omega_p$ is not resonant with the collective mode: we present the corresponding two-dimensional plot within the HF theory in Appendix C. However, this does not hold in the ED results, which may be due to photoinduced collective motions of CO with nonzero momenta \cite{Hashimoto_PRB2017} that are not captured in the HF approximation. 

\section{Summary}
In summary, we investigated the photoinduced melting process of CO depending on the light frequency $\omega_p$ using the interacting 
spinless fermion model in one dimension. We calculated time evolution of the system with pulsed laser light and transient spectral functions by two methods, the HF approximation and exact diagonalization, and the results are examined comparatively. We have shown that, as a common feature for the two methods, when the light frequency $\omega_p$ is nearly resonant with the collective mode, the in-gap weight appears in transient spectral functions and its development collapses the gap structure with increasing the laser strength. This is in sharp contrast to the case with $\omega_p>\Delta_{\rm CO}$ where the photoirradiation mainly induces interband excitations that weaken the CO leading to the shrinkage of the gap. We have also shown that, compared with the HF method, the exact diagonalization gives efficient destabilization of the CO in a wider region of $\omega_p$, which is due to the effect of quantum fluctuations.  

\begin{acknowledgments}
This work was supported by JSPS KAKENHI Grant Numbers 23K03309, 23K25826, 23K03333, and 25H00838. 
\end{acknowledgments}

\appendix
\section{Linear optical absorption and optical conductivity}
In this appendix, we compare our results of the linear optical absorption spectrum $\alpha(\omega)$ shown in Figs. \ref{fig1}(a) and \ref{fig5}(a) with optical conductivity 
spectra calculated within the linear response theory. Considering that the power absorption $P$ due to continuous electric field of light is estimated as $P\propto \sigma_{\rm reg}(\omega)E_{\rm cw}^2$ where $\sigma_{\rm reg}(\omega)$ and $E_{\rm cw}$ are the regular part of optical conductivity and the amplitude of the light electric field, respectively, $\alpha(\omega)$ is given as $\alpha(\omega)\simeq P\Delta T\propto \sigma_{\rm reg}(\omega)(\omega A_{\rm cw})^2 \Delta T$ with $\Delta T$ being the measurement duration for the increment in total energy. In the linear response theory, $\sigma_{\rm reg}(\omega)$ is calculated as
\begin{equation}
\sigma_{\rm reg}(\omega)=-\frac{1}{N\omega}\langle \Psi_0|j\frac{1}{\omega-\mathcal{H}+E_0+i\eta}j|\Psi_0\rangle,
\end{equation}
where $|\Psi_0\rangle$ is the ground-state wave function, $E_0$ the ground state energy, and $j$ the current operator that is defined as
\begin{equation}
j=-it\sum_{i}(c^{\dagger}_ic_{i+1}-c^{\dagger}_{i+1}c_i).
\end{equation}

\begin{figure}
\includegraphics[width=9.0cm]{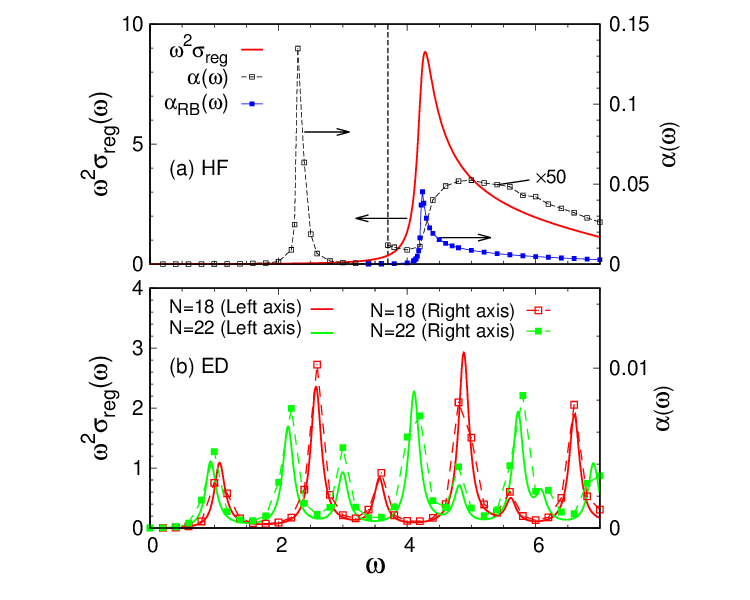}
\caption{(a) $\omega^2 \sigma_{\rm reg}(\omega)$ calculated within the HF theory, which is compared with $\alpha(\omega)$ shown in FIg. \ref{fig1}(a). We also show $\alpha_{\rm RB}(\omega)$ that indicates $\alpha(\omega)$ for the rigid band system. We note that $\omega^2 \sigma_{\rm reg}(\omega)$ and $\alpha_{\rm RB}(\omega)$ are depicted in a linear scale. (b) $\omega^2 \sigma_{\rm reg}(\omega)$ obtained by the ED method for $N=18$ and $22$, which are compared with  $\alpha(\omega)$ shown in Fig. \ref{fig5}(a). We use $V=3$ and $\eta=0.1$.}
\label{fig_A1}
\end{figure}

We first discuss the results obtained by the HF theory. In Fig. \ref{fig_A1}(a), we show $\omega^2 \sigma_{\rm reg}(\omega)$ for $V=3$ and $\eta =0.1$ where we depict $\alpha(\omega)$ in Fig. \ref{fig1}(a) for comparison. The quantity $\omega^2 \sigma_{\rm reg}(\omega)$ has a single peak at $\omega=\Delta_{\rm CO}$ corresponding to the CO gap in the ground state accompanied by a continuum-like tail above it, which is in contrast to $\alpha(\omega)$ that has a peak at $\omega=\Omega_c \sim \Delta_{\rm CO}/2$ due to the collective mode and the continuum above the CO gap. This discrepancy comes from the fact that in $\sigma_{\rm reg}(\omega)$ the HF order parameters are treated as static quantities, whereas $\alpha(\omega)$ calculated by the time-dependent Schr\"odinger equation takes account of their dynamics at the level of the random phase approximation by which the collective mode can be captured.

To see this point explicitly, we define $\alpha_{\rm RB}(\omega)$ as $\alpha(\omega)$ for a rigid band system where the HF order parameters are artificially fixed during the time evolution under the electric field in Eq. (\ref{eqn13}). We show the result of $\alpha_{\rm RB}(\omega)$ in Fig. \ref{fig_A1}(a) indicating that it has only a single peak at $\omega=\Delta_{\rm CO}$ and has no peak at $\omega=\Omega_c$, which is the same spectral feature as that of $\omega^2 \sigma_{\rm reg}(\omega)$.

Next, we examine the results with the ED method. In Fig. \ref{fig_A1}(b), we show $\omega^2\sigma_{\rm reg}(\omega)$ for $V=3$ and $\eta=0.1$ with $N=18$ and $22$ together with the corresponding results of $\alpha(\omega)$ presented in Fig. \ref{fig5}(a). It is apparent that, apart from an overall scale factor, the spectral feature of $\omega^2 \sigma_{\rm reg}(\omega)$ almost coincides with that of $\alpha(\omega)$. This demonstrates that $\omega^2\sigma_{\rm reg}(\omega)$ calculated exactly by using the many-body wave function essentially gives the linear absorption spectrum, which is in contrast to the case of the HF theory.

\section{ED results for the decrease in $\phi$ on $\{\omega_p, A_p\}$ plane with $\tau_p=10$}

\begin{figure}
\includegraphics[width=8.0cm]{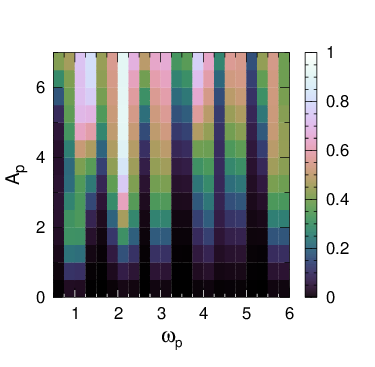}
\caption{Two-dimensional plot of $(\phi_0-|\phi_{\rm av}|)/\phi_0$ on the $\{\omega_p, A_p\}$ plane for $\tau_p=10$ obtained by many-body dynamics with the ED. We use $V=3$ and $N=22$.}
\label{fig_A2}
\end{figure}

In this Appendix, we discuss the effects of the pulse width $\tau_p$ on the decrease in $\phi$ due to photoexcitation. In 
Fig. \ref{fig_A2}, we show the two-dimensional plot of $(\phi_0-|\phi_{\rm av}|)/\phi_0$ on the $\{\omega_p, A_p\}$ plane for $\tau_p=10$ and $N=22$ obtained by the ED method. As we have discussed in Sec. III B, the structure of Fig. \ref{fig_A2} essentially reflects that of $\alpha(\omega)$ presented in Fig. \ref{fig5}(a). Compared to the case with $\tau_p=3$ shown in Fig. \ref{fig9}, each frequency range over which $\phi$ largely decreases becomes narrower. This is because the light with longer pulse width excites the system over a narrower frequency window. In particular, a large decrease in $\phi$ seen for the low-$\omega_p$ region ($\omega_p\simeq 0.5$) in Fig. \ref{fig9} comes from high frequency components contained in the short pulsed light with $\tau_p=3$ and is not due to an adiabatic response, which is indicated by the results with $\tau_p=10$ (Fig. \ref{fig_A2}) where no noticeable decrease in $\phi$ appears in that region.

\section{HF results for the decrease in $\phi$ on $\{\omega_p,A_p\}$ plane}

\begin{figure}
\includegraphics[width=8.0cm]{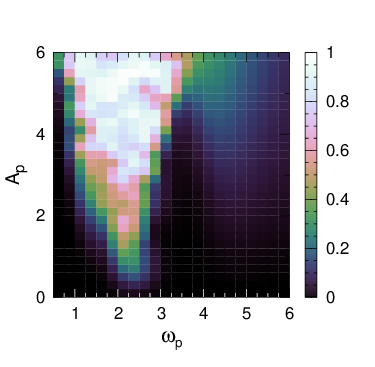}
\caption{Two-dimensional plot of $(\phi_0-|\phi_{\rm av}|)/\phi_0$ on the $\{\omega_p, A_p\}$ plane obtained by the time-dependent HF method. We use $V=3$, $\tau_p=3$, and $N=400$.}
\label{fig_A3}
\end{figure}

In Fig. \ref{fig_A3}, we show $(\phi_0-|\phi_{\rm av}|)/\phi_0$ on the $\{\omega_p,A_p\}$ plane obtained by the time-dependent HF method with $N=400$ where the other parameters are the same as those of the ED results shown in Fig. \ref{fig9}. When $\omega_p$ is resonant to the collective mode ($\omega_p= \Omega_c\simeq \Delta_{\rm CO}/2$), a large decrease in $\phi$ appears, whereas a relatively weak response appears around $\omega_p= \Delta_{\rm CO}$ where the interband excitations suppress the CO. This result is qualitatively the same as those obtained in our previous HF studies with different parameter sets~\cite{Seo_PRB2018,Seo_PRB2024} and can be understood from the frequency dependence of $\alpha(\omega)$ shown in Fig. \ref{fig1}(a).

%


\end{document}